\documentclass[a4paper,square,sort,comma,numbers,11pt]{article}
\usepackage{jheppub} 
\usepackage{upgreek} 
\usepackage{float}
\usepackage{graphicx}   
\usepackage{amsmath}
\usepackage{ulem}
\usepackage{textcomp}
\usepackage{import}
\usepackage{subcaption}
\usepackage{lineno}
\usepackage{hyperref}
\usepackage{url}
\usepackage{xcolor}
\usepackage[inkscapeformat=png]{svg}
\title{\boldmath Deployment and performance of a Low-Energy-Threshold Skipper-CCD inside a nuclear reactor}

\author[a,b,1]{E. Depaoli,\note{Corresponding author.}}
\author[b,c]{D. Rodrigues,}
\author[d]{I. Sidelnik,}
\author[a]{P.Bellino,}
\author[e]{A.Botti,}
\author[f]{D. Delgado,}
\author[b,c,e]{M.Cababié,}
\author[g]{F.Chierchie,}
\author[e]{J.Estrada,}
\author[e]{G.Fernández Moroni,}
\author[b,c,e]{S.Perez,}
\author[e]{J.Tiffenberg.}

\affiliation[a]{\normalsize\it Comisión Nacional de Energía Atómica, Centro Atómico Constituyentes}
\affiliation[b]{\normalsize\it Universidad de Buenos Aires, Facultad de Ciencias Exactas y Naturales, Departamento de Física, Buenos Aires, Argentina}
\affiliation[c]{\normalsize\it CONICET - Universidad de Buenos Aires, Instituto de Física de Buenos Aires (IFIBA). Buenos Aires, Argentina}
\affiliation[d]{\normalsize\it Departamento de Física de Neutrones, Centro Atómico Bariloche, (CNEA, CONICET), Bariloche, Argentina}
\affiliation[e]{\normalsize\it 
Fermi National Accelerator Laboratory, PO Box 500, Batavia IL, 60510, USA}
\affiliation[f]{\normalsize\it Central Nuclear ATUCHA I-II, Nucleoeléctrica Argentina Sociedad Anónima, Buenos Aires, Argentina.}
\affiliation[g]{Instituto de Inv. en Ing. El\'ectrica ``Alfredo Desages'' (IIIE), CONICET and Universidad Nacional del Sur (UNS), Bah\'ia Blanca, Argentina.}

\emailAdd{edepaoli@df.uba.ar}

\abstract{Charge Coupled Devices (CCD) are being used for reactor neutrino experiments and have already demonstrated their potential in constraining new physics models. The prospect of a Skipper-CCD experiment looking for standard and beyond standard model (BSM) physics in a nuclear reactor has been evaluated for different benchmark scenarios. 
Here we report the first installation of the 2 g Skipper-CCD inside the containment building a 2 GW$_{th}$ nuclear power plant, 12 meters away from the center of the reactor core. 
We discuss the challenges involved in the commissioning of the detector and present data acquired during reactor ON and reactor OFF periods, with the detector operating with a sub-electron readout noise of 0.17 e$^-$. The ongoing efforts to improve sensitivities to CEvNS and BSM interaction are also discussed.}

\begin{document} 

\maketitle
\flushbottom
\label{sec:introduction}

\section{Introduction}

The observation of Coherent Elastic Neutrino Nucleus Scattering (CEvNS)~\cite{PhysRevD.9.1389} by COHERENT collaboration using an accelerator beam~\cite{COHERENT:2017ipa} encouraged the interest in observing this interaction in nuclear power reactors, the Earth's most powerful sources of neutrinos. Despite the significant increase in the number of short-baseline reactor neutrino experiments in recent years~\cite{CONNIE2022,CONUS:2020skt,PROSPECT:2018dnc,TEXONO:2018nir,Strauss:2017cuu,Majumdar:2022nby,nGeN:2022uje,NEON:2022hbk,Akimov:2022xvr,SoLid:2020cen,MINER:2016igy}, none of them has achieved the required sensitivity to observe CEvNS yet. However, they proved to be competitive in imposing constraints on Beyond the Standard Model (BSM) scenarios that predict a significant number of events at low energies~\cite{Aguilar_Arevalo_2020, CONUS:2022qbb, CONUS:2021dwh}.

In light of the success of the Skipper-CCD technology used for Dark Matter searches~\cite{SENSEI2020}, its extension to low-energy neutrino physics becomes natural. Lowering the energy threshold provides significant benefits, as the cross-section of many BSM interactions significantly increases with decreasing energy. Thus, the sensitivity of short-baseline experiments using the Skipper-CCD technology was studied. A competitive precision on weak mixing angle measurements via CEvNS was evaluated~\cite{VIOLETA2021} in addition to sensitivity to non-standard interactions of neutrinos mediated by light particles~\cite{VIOLETA2022}.
Since exploring the CEvNS channel requires good background control, the background shape measured by a Skipper-CCD at ground level was previously studied. As a result, a uniform spectrum down to 5 electrons was observed, in tension with the excess observed by other low-energy experiments~\cite{Moroni2022}.

\begin{figure}
    \centering
    \includegraphics[width=0.9\textwidth]{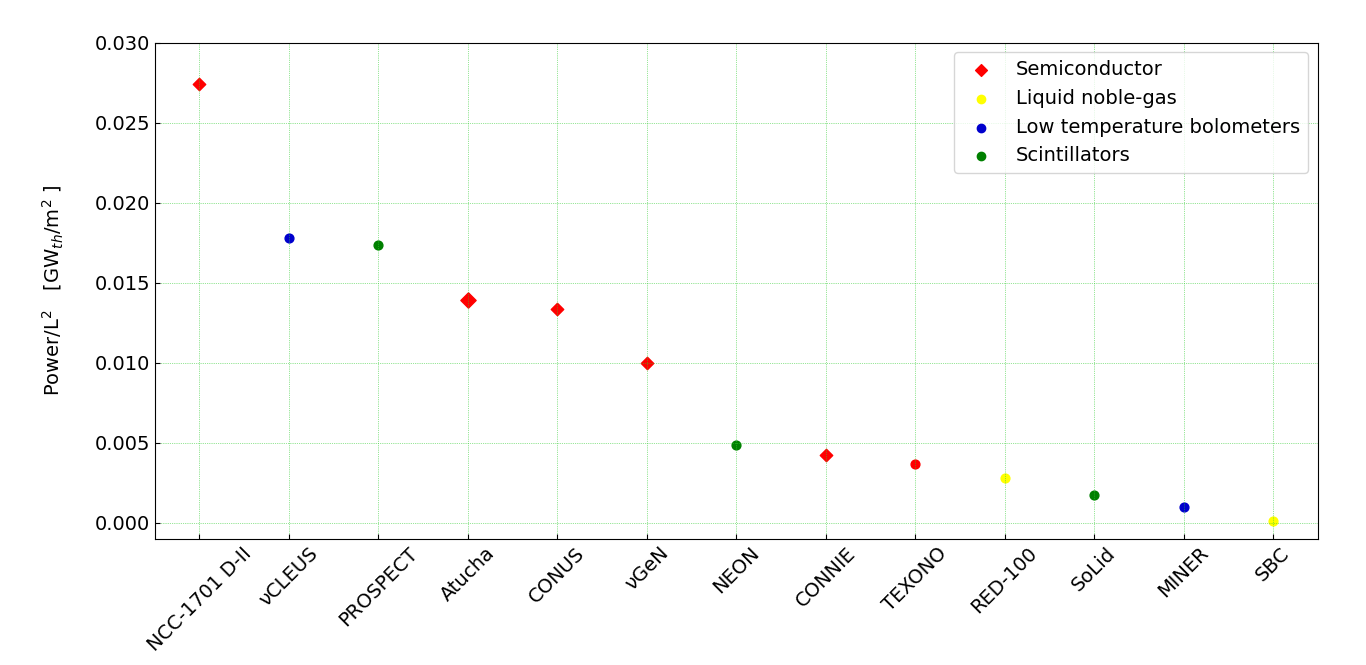}
    \caption{The ratio between the thermal power of the nuclear reactor and minimum distance (L) to the central core for each short-baseline neutrino experiment. Data extracted from~\cite{CONNIE2022,CONUS:2020skt,PROSPECT:2018dnc,TEXONO:2018nir,Strauss:2017cuu,Majumdar:2022nby,nGeN:2022uje,NEON:2022hbk,Akimov:2022xvr,SoLid:2020cen,MINER:2016igy} and summarized in Ref.~\cite{AristizabalSierra:2020rom}.}
    \label{fig:cevns_experiment}
\end{figure}

A parameter of interest in short baseline neutrino experiments is the ratio of the reactor power to the square of the distance between the core and the detector, as the antineutrino flux is proportional to it. 
Figure~\ref{fig:cevns_experiment} presents a comparison of that ratio for several reactor neutrino experiments with sub-keV detection thresholds. Atucha corresponds to the experiment described in this work, where the first Skipper-CCD experiment running inside a nuclear reactor is presented, located just 12 meters from the center of the core. 

We first describe the nuclear power plant, the background studies conducted before sensor deployment, the Skipper-CCD sensor, and the equipment associated with it. As our early-stage outcomes, we detail the current data collection performance and data processing and introduce reactor spectra under two different conditions. Based on these results, the upcoming steps and prospects for this experiment are also discussed.

\section{The nuclear power plant}
\label{sec:Site}

The Nuclear Power Plant \textit{Atucha II}, located in the Buenos Aires Province of Argentina and operated by \textit{Nucleoelectrica Argentina S.A} (NA-SA), has a thermal power of  2\,GW$_{th}$ (745\,MWe) and began delivering energy to the Argentinian electrical system in 2014. It is a pressurized heavy water reactor of German design (Siemens KWU) containing 451 fuel elements with UO$_2$ pellets. Each fuel element is cylindrical, with an active length equal to 5.3\,m and a diameter of nearly 10\,cm. The fuel elements are vertically allocated inside the pressure vessel (14\,m high and 8.44\,m wide) in a hexagonal grid with a center-to-center distance of 27.2\,cm.

\begin{figure}
    \centering
     \begin{subfigure}[b]{0.5\textwidth}
        \centering
        \includegraphics[width=.95\textwidth]{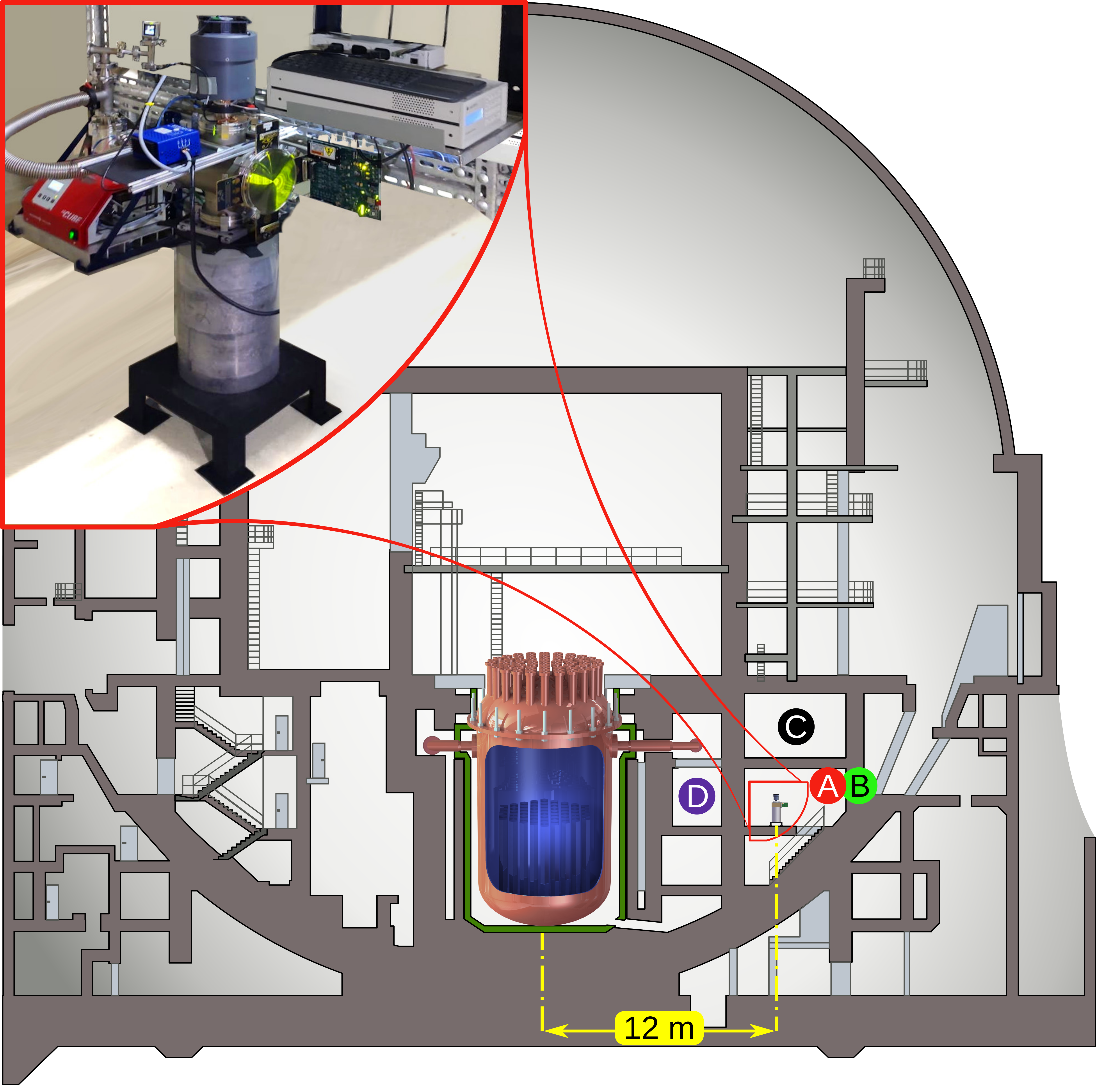}
    \end{subfigure}
     \begin{subfigure}[b]{0.48\textwidth}
        \centering
        \includegraphics[width=.55\textwidth, trim= 0 0 0 0,clip ]{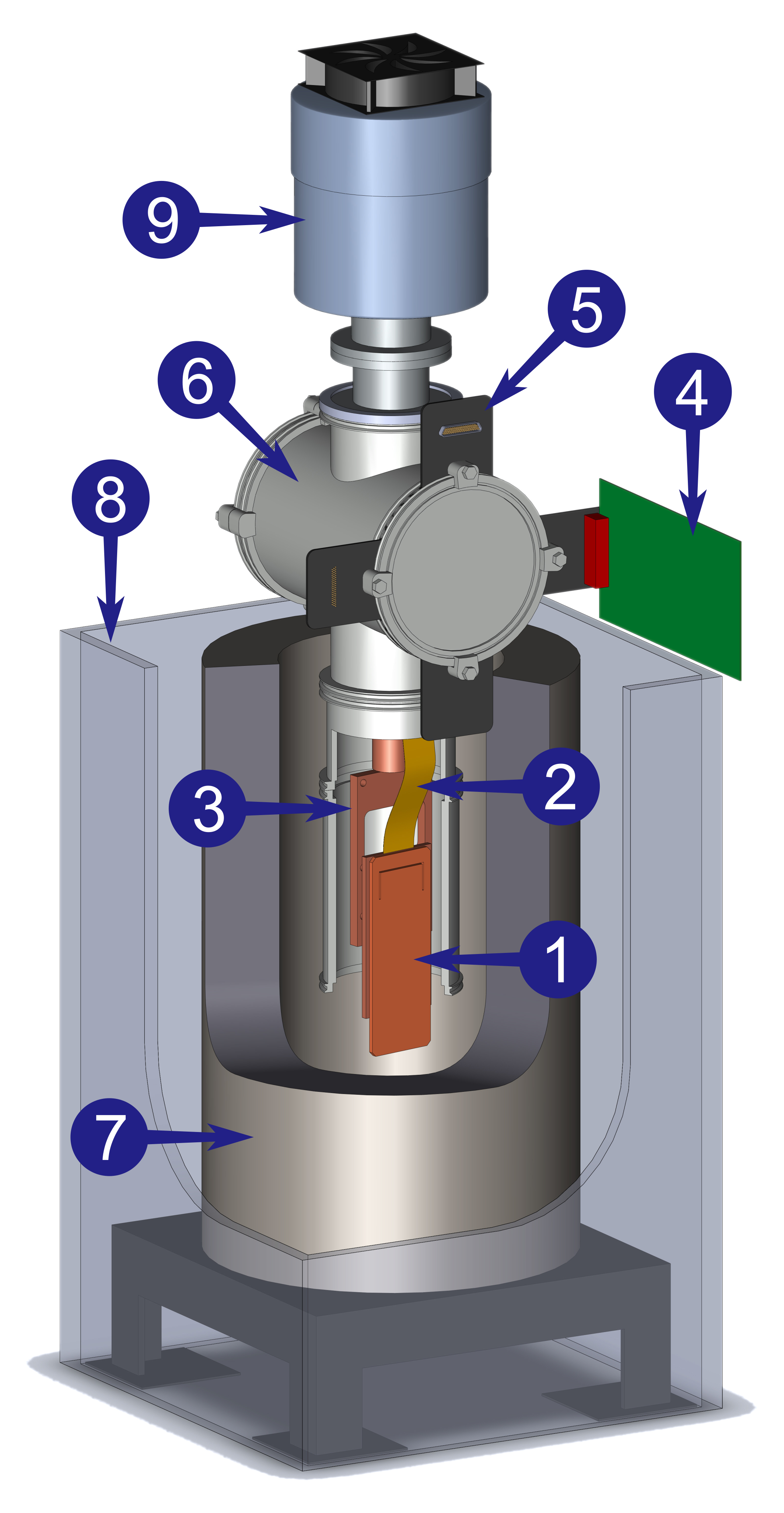}
    \end{subfigure}
    \caption{View of the containment sphere of Atucha II. The Skipper-CCD was installed in Site A. Sites B, C, and D were dismissed due to comparatively higher background. \textbf{Right.} Most remarkable features of the experiment: 1. Skipper-CCD inside copper case, 2. Flex cable, 3. Copper tray, 4. LTA, 5. VIB, 6. Dewar, 7. Lead shield, 8. Polyethylene shield, 9. Cryocooler.}
    \label{fig:system-in-Atucha}
\end{figure}

In 2019, we started conversations to evaluate the feasibility of implementing a neutrino experiment in this facility. As we explored the technical requirements, four sites emerged as viable options. These locations are highlighted in the sketch of the containment sphere of Atucha II on the left panel of figure~\ref{fig:system-in-Atucha}, illustrating both the vessel and the reactor core, along with the final layout of the Skipper-CCD. The current position of the device (Site A) is shielded from most of the background radiation coming from the reactor by a double concrete wall 3 meters wide. The dose measured at this position is around 1\,$\mu$Sv/h, the same value as outside the reactor at ground level. 

\section{Background studies}\label{sec:bkg}

\subsection{Gammas}

\begin{figure}[t]
\centering
\includegraphics[width=.9\textwidth]{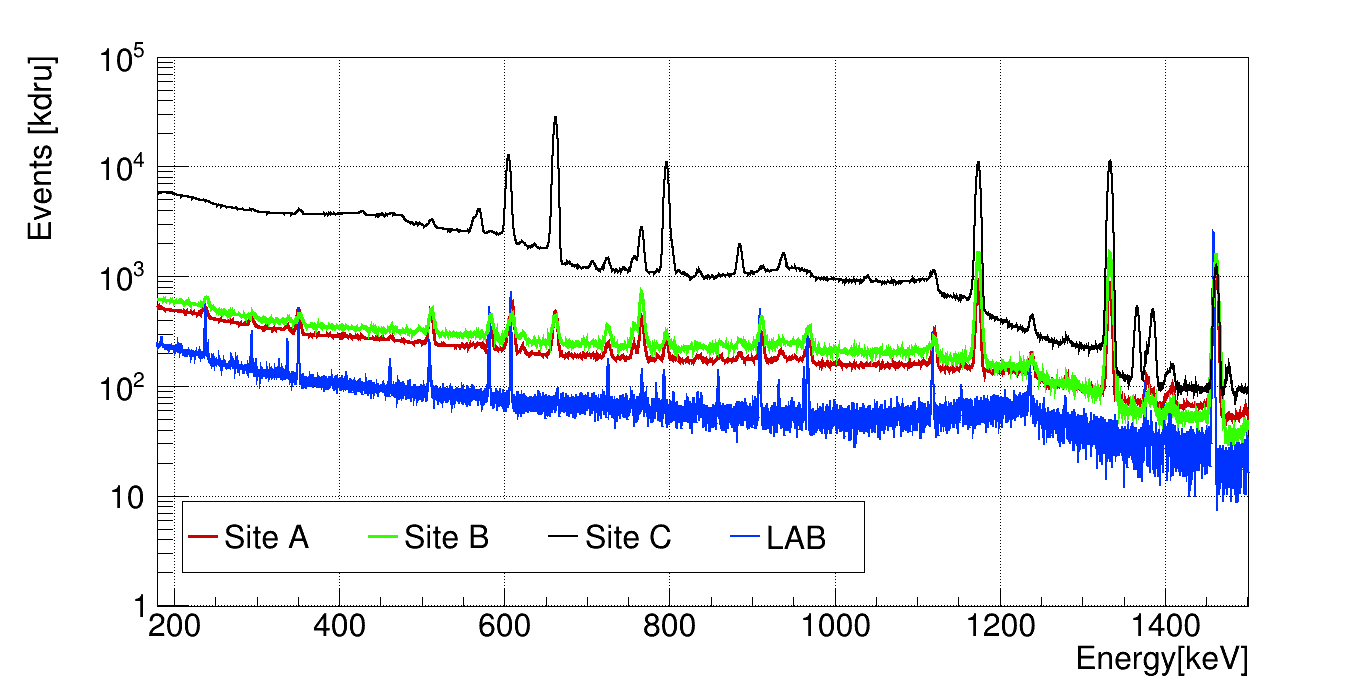}
\caption{Gamma background measured inside the containment sphere of Atucha II, all sites situated approximately $12$ m away from the reactor core and at laboratory conditions, 100 km from the nuclear plant. Site A is the current position of the Skipper-CCD.}
\label{fig:gamma_spectra_dome_caba}
\end{figure}

Exploratory gamma background measurements were conducted in Site A, B and C inside the containment building, all three located 12 meters away from the core center. Sites A and C are depicted in the illustration (figure~\ref{fig:system-in-Atucha}. Left panel), while Site B is situated adjacent to A at the same level. In addition, we perform a reference measurement in laboratory conditions 100 km away from the power plant. With that goal, we employed a portable HPGe detector consisting of one crystal of mass 495 g maintained at 77 K using a Stirling-activated water vapor pump. This detector has a resolution of 3.29 keV at the 1332.5 keV full energy peak of Co-60.

Acquired spectra are shown in the range 0.15 to 1.50 MeV in figure~\ref{fig:gamma_spectra_dome_caba}. Photo peaks from heavy metals such as Pb-212, Pb-214, Tl-208, Bi-214, and K-40 appear both inside and outside the containment building. Nevertheless, the Compton continuum from the high-energy gamma rays in the laboratory accounts for a threefold lower event rate than the best spectrum acquired inside the power plant. Furthermore, it is observed that the gamma spectrum integral inside the dome is site-dependent. As a result, we decided to set up the Skipper-CCD in Site A where we observed the lowest gamma background.

Cumulative counting rates were calculated over ten hours of measurements in Site A for both reactor ON and OFF. The difference between these two conditions, from 65 keV up to 600 keV, was found to be smaller than 1\% and considered negligible. As a consequence, we do not expect a significant increase of gamma background in this range of energies due to fuel burnt when the reactor is operative.

Site D, (see figure~\ref{fig:system-in-Atucha}. Left), was dismissed because it is accessible only during reactor OFF periods and the measured dose in there was considered capable of producing undesirable effects on the electronic items of the Skipper-CCD setup.

\begin{figure}[t]
\centering 
\includegraphics[width=.90\textwidth]{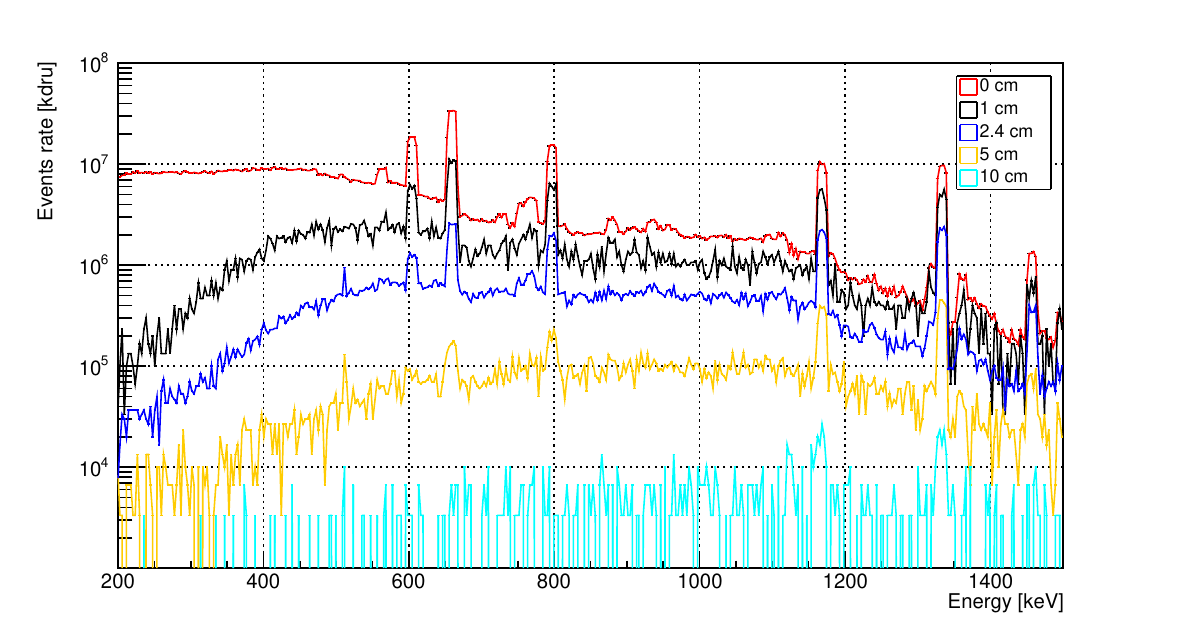}
\caption{\label{fig:gammas_MC} Propagation of gamma background measured in site A (red) through lead of varying thickness using PENELOPE Monte Carlo simulation code~\cite{penelope}.}
\end{figure}

With the end of quantifying the effect of adding a shield around the sensor, the propagation of gamma background measured in Site A through lead of varying thickness was simulated using PENELOPE Monte Carlo code~\cite{penelope}. 
As shown in figure~\ref{fig:gammas_MC}, 5~cm of lead would reduce by at least one order of magnitude the rate of photons arriving at the CCD with energies lower than 1 MeV. Moreover, 10 cm of lead would decrease even the gamma rays produced by Co-60 decay by almost two orders of magnitude.

\subsection{Neutrons}
\label{subsection:neutrons}
\begin{figure}[t]
    \centering
    \includegraphics[width=0.84\textwidth]{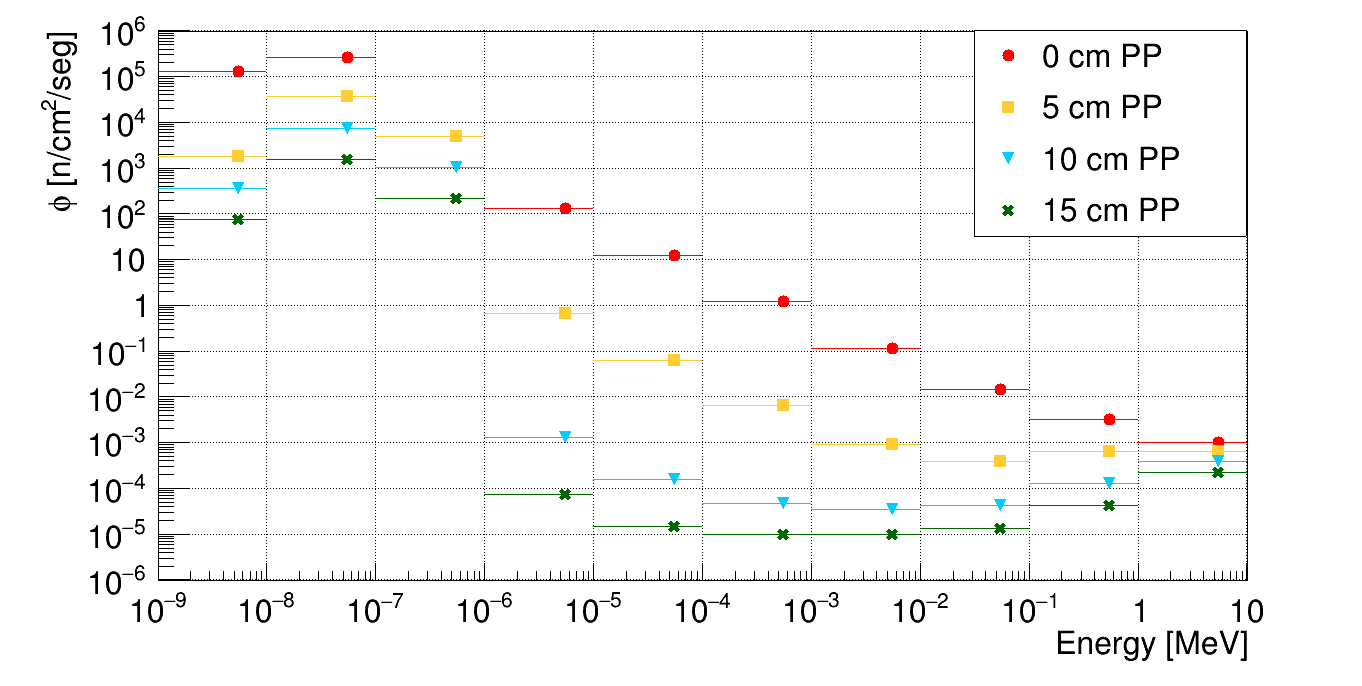}
    \caption{ Simulated neutron flux produced by fuel burnt reaching Site A (red) after passing through a 3-meter-thick concrete wall. This flux transported through various thicknesses of polypropylene (PP) is also presented.}
    \label{fig:neutron_simulation}
\end{figure}

The neutron flux expected to reach Site A, where the Skipper-CCD setup is placed, was obtained through various stages of Monte Carlo simulation. Firstly, the rate of flow leaving the reactor vessel was simulated by NA-SA team \cite{cnainternal}. Then, we transported the three most energetic groups of the resulting flux through 
3 meters of concrete 
thickness to emulate the shielding between the reactor vessel and Site A using the PHITS 3.17~\cite{Phits} Monte Carlo code. The ``Cold to Thermal'' set was omitted from this calculation, anticipating that it would be captured by the concrete. Also, a weight window scheme was used as a variance reduction technique to shorten the computation time. 
As a result, depicted in the red spectrum of figure~\ref{fig:neutron_simulation}, we should expect that the majority of the neutrons arriving at the Skipper-CCD device have an energy lower than 1 keV, which could be stopped with a few centimeters of a hydrogen-rich material as polypropylene. To quantify the effect of this sort of shield, we transported this flux through various thicknesses of polypropylene. 
According to the resulting spectra, also shown in figure~\ref{fig:neutron_simulation}, with just 5 cm of thickness, the incident flux is strongly thermalized and the total flux is reduced $\sim$ 89 $\%$. Furthermore, 10 cm of polypropylene is capable of virtually removing the intermediate and rapid components of the spectrum, which could otherwise leave a signature in the Skipper-CCD, and reduces the total neutron flux in $\sim$ 98 $\%$.
It is important to note that, due to the kinematics of neutron scattering with silicon nuclei, at most, one-seventh of their energy is transferred. Then, only a fraction of up to 10\% is used for ionization due to the quenching factor. Consequently, neutrons of 1 keV will produce less than 15 eVee, below the minimum detection threshold of the system.

\section{Detector features}
\label{sec:sys_deployment}
Skipper-CCDs are imaging sensors with sub-electron readout noise achieved through multiple nondestructive readouts of each charge packet. This capability along with the 1.1~eV band gap energy in silicon made this technology especially suitable for sensing very weak ionizing particles.

The Skipper-CCD was installed at Atucha II in December 2021 and has been acquiring data since then. The setup is depicted on the right panel of figure~\ref{fig:system-in-Atucha}. The sensor, designed by LBNL Microsystems Laboratory and fabricated at Teledyne-DALSA, has a spatial resolution of 6.29 MPx, with a pixel size of 15~$\mu$m $\times$ 15~$\mu$m, a total mass of 2.22 grams, an active area of 9.216 $\times$ 1.536 cm$^{2}$ and a thickness of 675~$\mu$m. 
This science-grade Skipper-CCD is placed inside a copper case which shields it from the infrared radiation from the vessel's walls.
The case is thermally coupled to the cryocooler cold tip using a copper cold finger wrapped in multi-layer insulation used to reflect black body radiation from surroundings. A stainless steel piece exerts constant pressure between the cold finger extension and the cold tip to improve thermal contact.
The ensemble is vertically located in a T-shaped dewar which is lodged inside a 5~cm thick lead cylinder. The dewar, the pump station, and the cryocooler are electrically grounded in a star configuration to reduce noise loops, and their paths to the ground are isolated from the CCD line to avoid low-frequency noise from being injected into the readout system. Finally, the setup is placed over a support which provides insulation from mechanical vibrations. To prevent humidity deposition over the sensor, the air is evacuated from dewar employing a turbo pumping station. The system operates at 130~K using a Sunpower CryoTel \textsuperscript{\textregistered} GT cryocooler and is maintained under a vacuum pressure of approximately ~10$^{-5}$ mbar. The readout is performed through a Low-threshold Acquisition (LTA) board~\cite{LTA2021}.

The system's performance at ground level in laboratory conditions was previously evaluated at Fermilab \cite{Moroni2022}. It was then disassembled to be transported to Argentina and reassembled at the nuclear power plant in a laboratory outside the containment sphere. Finally, it was transported to its final position inside the reactor, from where it is remotely monitored via a VPN connection. 

\section{Data acquisition}
\label{sec:Data_acquisition}

The Skipper-CCD has four amplifiers, one in each corner, capable of synchronously reading the charge in the quadrant to which they belong, thereby producing four subsets of images. The Skipper-CCD was configured to run with a hardware horizontal binning in each quadrant so that 10 consecutive pixels in the same row are added and treated as one. Two of the four quadrants work in expected conditions, achieving a sub-electron readout noise of $\sigma_{\text{RN}}$=0.17 e$^{-}$/pix by averaging $N$=300 samples of the charge in each pixel~\cite{Tiffenberg2017}. The readout time required for each full image is 53 minutes. 

The measurements were performed under two different acquisition modes, outlined in Table~\ref{tab:datasets} as DATA SET A and B. The former was collected in continuous readout mode, where the active array of the sensor is continuously read, resulting in the same exposure for all the pixels. For this group of images, the setup had a 5~cm lead shield. For DATA-SET B, a fast single readout of the entire CCD was conducted as a cleaning routine between images. This adjustment alters the effective exposure time of each pixel due to the readout duration, consequently halving the integral exposure time for a full image. Additionally, 24 kg of high-density polyethylene was introduced as a neutron shield.

The DATA SET 0 corresponds to measurements performed with the system running at Fermilab for approximately 3.3 days and they were used to characterize the performance of Skipper-CCD for Low-Energy-Threshold Particle Experiments above ground~\cite{Moroni2022}.

\begin{table}
\centering
\caption{Data sets used. SET 0 lab conditions, while SET A and SET B were acquired inside the reactor. SET A utilized continuous readout mode, whereas for SET B, a single sample per pixel image was taken to clean the CCD between science images. An additional polyethylene shield was introduced to the setup for measurements in SET B.}
\label{tab:datasets}
\begin{tabular}{||c c c c||} 
 \hline
 DATA & Reactor & Exposure   & Shield \\ [0.5ex] 
  SET &  Status & [ g . day] &      \\
 \hline \hline
 0 & -- &  1.6 & Lead \\
 \hline
 A & OFF &  95.5 & Lead \\
 A & ON &  56.8  & Lead\\ 
 \hline
 B & OFF &  22.6 & Lead + poly \\
 B & ON &  running & Lead + poly \\
 \hline
\end{tabular}
\end{table}

\section{Data Analysis}
\label{sec:data_analysis}

\subsection{Images processing}  
\label{set:processing}
The protocol used to process the images begins with averaging the $N$ samples and subtracting the baseline calculated for each row using an overscan~\cite{janesick}. The following stage is correcting the crosstalk between the signals read synchronously among different quadrants. This effect leads to a spurious occupancy in originally empty pixels arising from the coupling of the amplifiers reading out each quadrant. Consequently, it induces a linear relationship between the measured charge in a specific quadrant and that measured in the others. Although this effect is of order $\sim 10^{-4}$, its correction is crucial for low-energy searches.

Because Skipper-CCDs have a sub-electronic resolution, no dedicated measurement is needed to calibrate the sensor, instead, we used the Analog to Digital Units (ADU) of the charge histogram between empty and one electron pixels. Then, the 3.75 eVee for the electron-hole pair creation energy is used to translate ionized electrons in deposited energy \cite{Rodrigues2023}. 
To fine-tune the calibration within the keV range, the 8.048 keVee from X-rays produced by the fluorescence in the copper tray was also employed, yielding a correction of less than 2\%.
Non-linearities on this calibration method in a Skipper-CCD were studied in \cite{Rodrigues2021, Lapi2021}, where it is shown that small deviations from linearity can be neglected for the operation mode employed in this work.

Finally, adjacent and diagonal non-empty pixels are grouped into events, with a charge corresponding to the sum of each pixel charge.
A pixel is determined to contain $n$ electrons if its measured charge is found to be within the interval $[n-0.5, n+0.5)$. Hence, a readout noise value of 0.17 e$^{-}$/pix leads to a probability of misclassification of $8 \times 10^{-4}$.

\begin{figure}[t]
\centering
\includegraphics[width=1.0\textwidth]%
{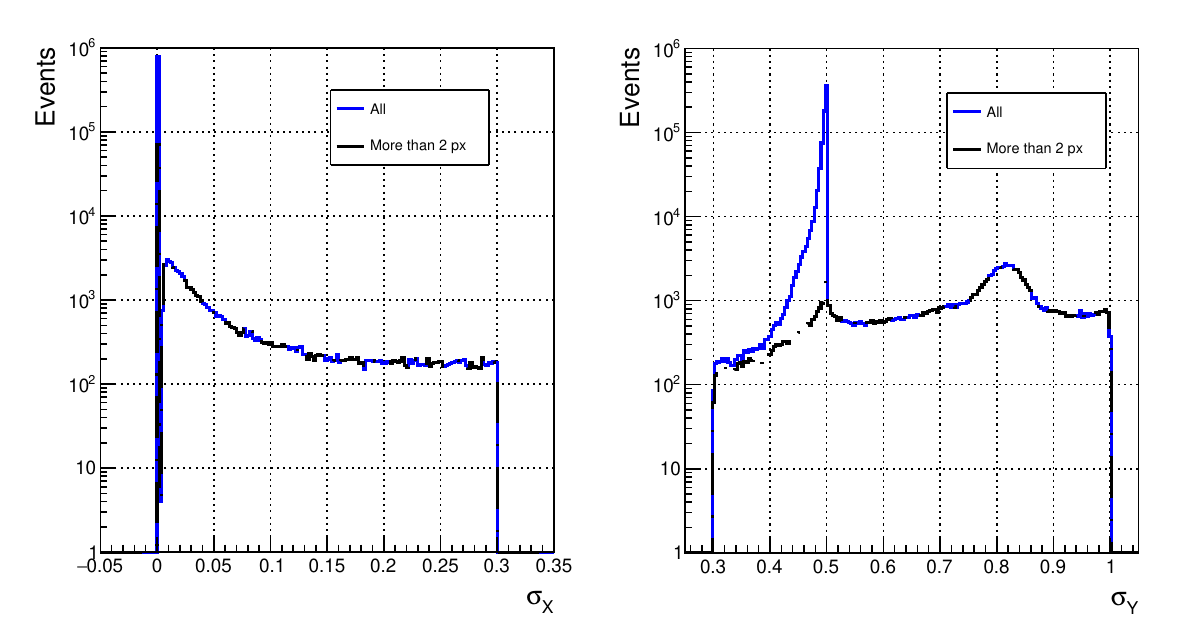}
\caption{Event size distribution after applying quality cuts (blue line). $\sigma_{x}$ and $\sigma_{y}$ are the standard deviation of the charge distribution in the pixels for an event along the x-direction and y-direction, respectively. The effect of excluding the one and two-pixel events is also shown (dashed black line). Plots produced using DATA SET A with reactor OFF.}
\label{fig:sigmas}
\end{figure}

\subsection{Quality cuts}

To ensure that the collected data meets quality criteria for the following analyses and spectra computation, the following quality cuts were applied.

\noindent \textbf{Edges of the active area}: a rectangular exclusion ring of 3 pixels was defined from the edge of the active volume to prevent counting clusters that have undergone geometric truncation. The criteria used here is to reject clusters whose charge barycenter lies within the exclusion region. As a result, the sensor's effective mass is reduced by 2.8~\%. 

\noindent  \textbf{Hot columns exclusion:}  hot columns are generated by defects in the silicon which manifests as illuminated straight vertical lines in the image. This leads to inefficiencies in charge transfer along the y-direction towards the readout stage. These defects are also responsible for an excess in the dark current~\cite{Oscura:2023qik}. Accordingly, columns with a value of the single electron events (SEE) rate higher than the remaining $90\%$ of the columns were classified as hot columns. 
Subsequently, clusters were excluded if any pixel within them belonged to a hot column.
 \\
\noindent \textbf{Geometry of events}: The charge distribution in every cluster was used to identify those produced within the sensor active volume \cite{Moroni:2021lyt}. The upper limit for the standard deviation along y-direction, $\sigma_{y} < 1.0 $,  is imposed by the maximum diffusion expected for a 675~$\mu$m width CCD \cite{SENSEI2020} while lower constraints along both x, y-direction, $0.3 < \sigma_{y}$ and $0 \leq \sigma_{x}$, are imposed to reject dark current. The latter along with the condition $\sigma_{x} < 0.3 $ aims to reject serial register events, the ones captured in the horizontal register (used to shift the charge sequentially into the sense node) and not in the active area.

\begin{figure}[t]
\centering 
\includegraphics[width=0.8\textwidth]{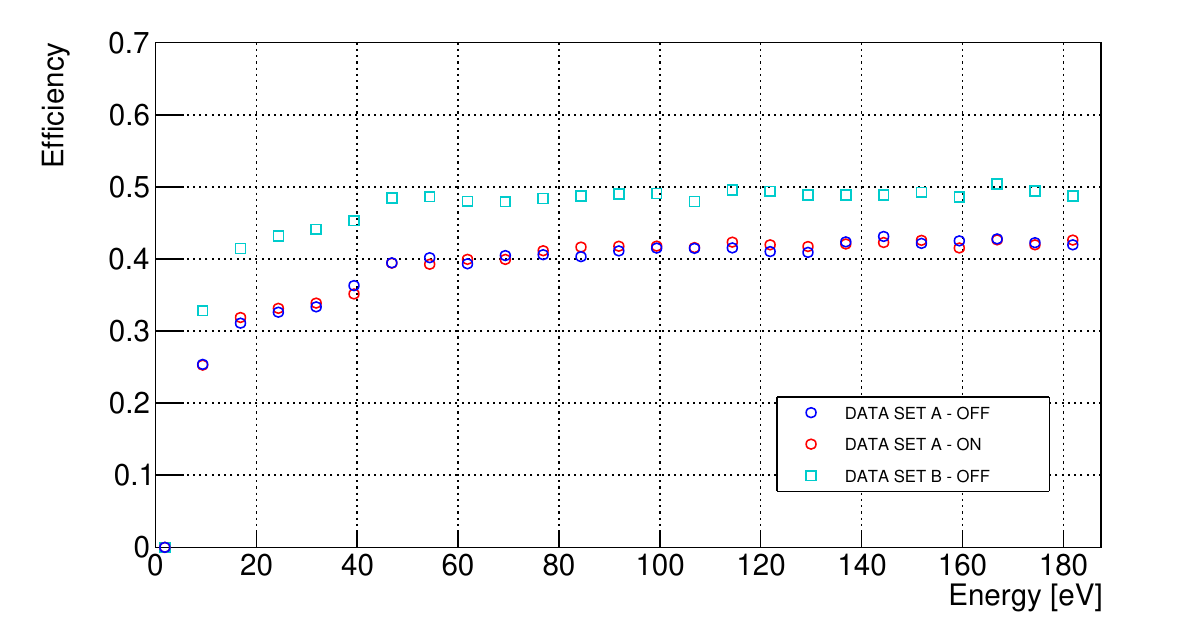}
\caption{Reconstructed efficiency as a function of the event's energy for DATA SET A (OFF and ON) and DATA SET B (OFF). 
\label{fig:eff}} 
\end{figure}

\subsection{Efficiency calculation} 
\label{set:efficiency}

To quantify the impact of the quality cuts on the reconstruction efficiency we implemented a toy-Monte-Carlo simulation based on a diffusion model~\cite{SENSEI2020} that describes how the electrons produced in the silicon bulk propagate towards the CCD surface. We injected events with 1 to 50 electrons in real images and computed the probability of successfully reconstructing the artificial cluster. This allowed us to obtain the reconstruction efficiency as a function of the number of electrons, which can be converted in energy as presented in figure~\ref{fig:eff} for DATA-SET in Table~\ref{tab:datasets}. 

If the interaction is produced close to the CCD surface, electrons will not spread much resulting in small clusters, while if the interaction happens in the back of the bulk the spread will be large and produce bigger clusters. In this sense, variance cuts indirectly constrain events to the bulk region where electrons are produced. However, at low energy, the maximum spread may strongly depend on the number of electrons regardless of where the interaction happens. Therefore, the selection efficiency depends on the interaction energy and distorts the spectrum shape. In addition, images with cleaning have lower occupancy than those without cleaning, which translates to a smaller probability of pile-up and, in turn,  gives a higher reconstruction efficiency. We obtained a flat efficiency for clusters with more than 12 electrons (45~eV) of 0.49 for DATA-SET A and 0.42 for DATA-SET B. We then applied an efficiency correction based on this result and normalized each dataset by the total exposure.

\section{Results}
\label{sec:results}

\begin{figure}[t]
\centering 
\includegraphics[width=.85\textwidth]{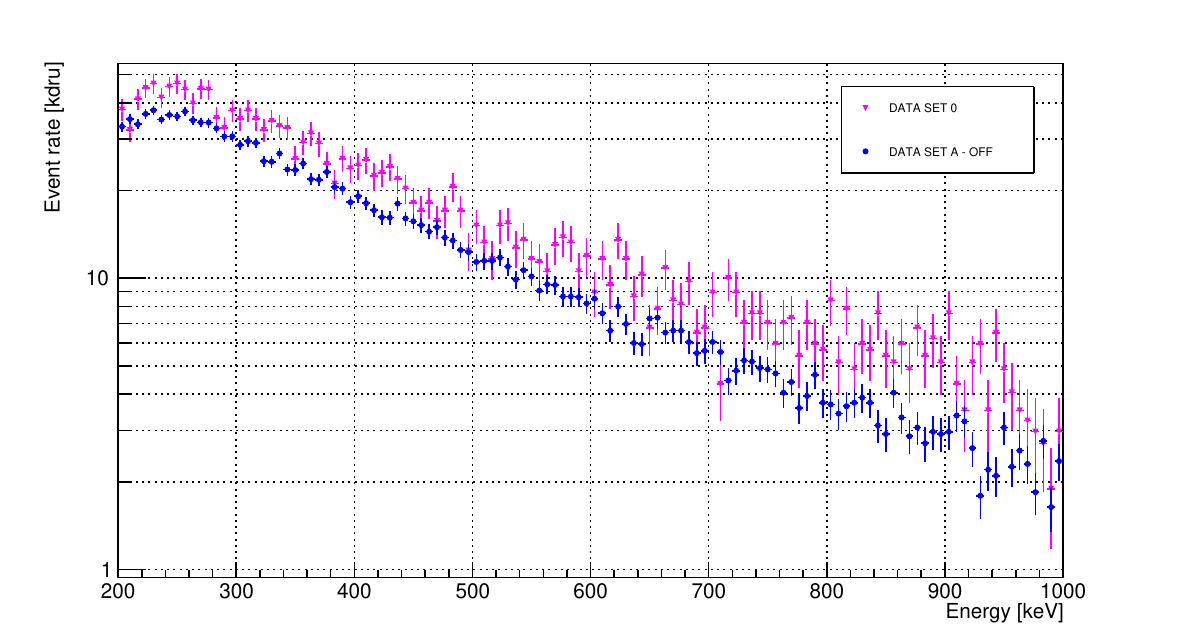}
\caption{Comparison of spectra for high-energy events at Fermilab and inside the reactor containment sphere when the reactor is off.}
\label{fig:spectraON2022OFF2023} 
\end{figure}

Figure~\ref{fig:spectraON2022OFF2023} compares the energy spectra ranging from 200~keV to 1000~keV
when running the detector at two different locations. This energy interval corresponds to the ionization typically generated by high-energy interactions from gamma and cosmic radiation. The magenta spectrum corresponds to DATA SET 0, measured when the system was running at Fermilab~\cite{Moroni2022}, while the blue distribution corresponds to a subset of images from DATA SET A OFF for 17.5 days. The rate of events inside the dome of Atucha is lower than at Fermilab, suggesting that the containment building of the nuclear power plant provides an overburden for cosmic radiation.
The total deposited energy of these events, normalized by the corresponding exposure times, is approximately 34~\% larger at Fermilab than inside the reactor containment building.

\begin{figure}[t]
\centering 
\includegraphics[width=0.9\textwidth]
{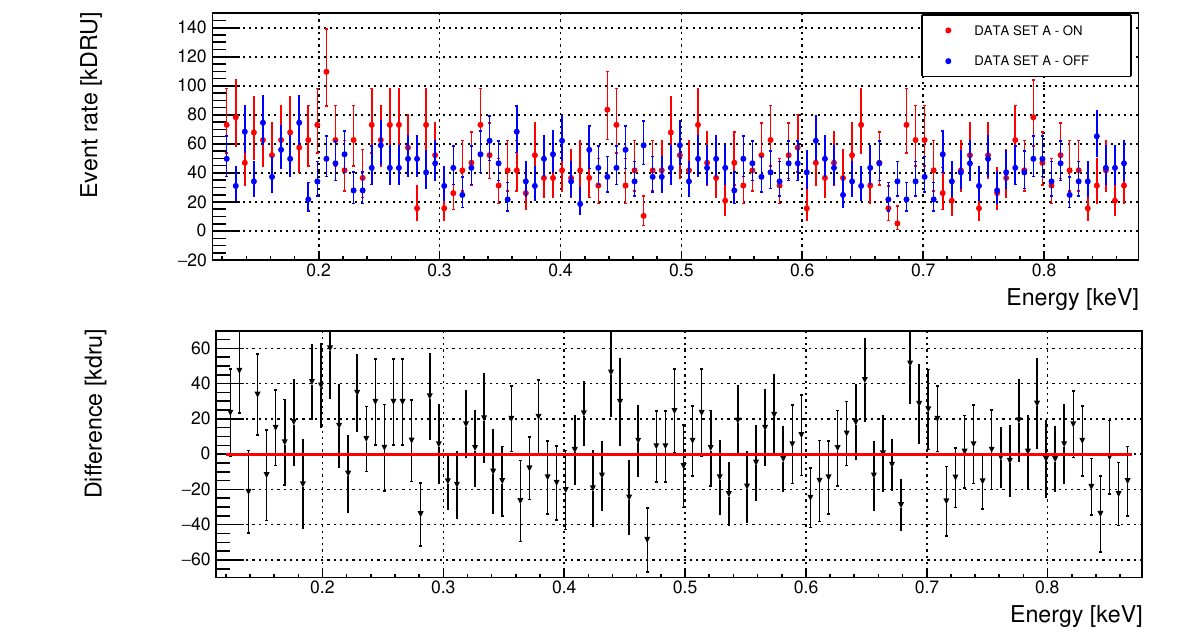}
\caption{Reactor ON and Reactor OFF spectra obtained from DATA SET A (continuous readout mode without neutron shield) ranging between 120 to 870 eV, top, along with their respective difference, show no significant deviation from zero (red line), bottom.}
\label{fig:spectraON2022OFF2022}
\end{figure}

Figure \ref{fig:spectraON2022OFF2022}, top, compares spectra obtained from DATA SET A during reactor ON and OFF, ranging from 120~eV up to 870~eV (200~e$^-$) with a bin size of 7.5~eV (2~e$^-$). The normalization of the spectrum is performed considering the sensor's active mass and the efficiency of the selection criteria. As verified from the reactor ON minus OFF spectrum on the bottom plot, the system provided with 5~cm of lead as shielding and situated $\sim~8$~m away from the core wall, does not see a significant difference between both conditions.  

\begin{figure}[t]
\centering
\includegraphics[width=0.9\textwidth]{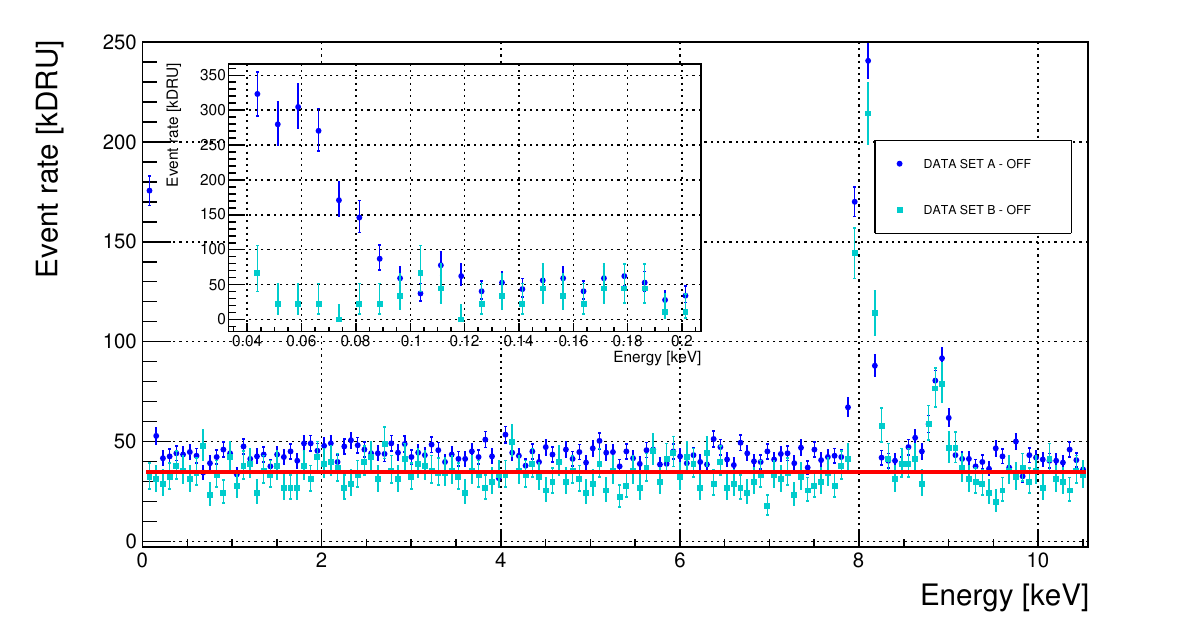}
\caption{\label{fig:spectraOFF2022OFF2023} Spectra under two different data acquisition modes when the reactor is OFF. DATA SET A corresponds to continuous readout mode, while during DATA SET B science images are alternated with clean images reducing the exposure by half. The red line represents a mean background rate of 35~kdru without including the fluorescence peak.}
\end{figure}

Figure \ref{fig:spectraOFF2022OFF2023} compares the reactor OFF spectra from DATA SET A and B with a bin size of 75~eV (20~e$^-$). The peaks corresponding to the fluorescence X-ray photons emitted by Cu, both the K$_{\alpha}$ K$_{\beta}$ lines, are visualized.
The spectrum from DATA SET B remains constant across the entire range below $\sim$8~keV. The mean rate of events, between 45 to 10545~eV (12 to 2800~e$^-$, where the efficiency is constant), corresponds to $\sim$35~kdru (red line).

The inset magnifies the region between 45 and 200~eV with a bin size of 7.5~eV (2~e$^-$) to highlight the impact of the sensor routine cleaning between images during the acquisition of DATA SET B. 
This significant reduction of background below 100 eV, now appearing flat down to 45 eV, is explained in terms of pile-up mitigation due to the high density of events populating the image acquired in continuous readout mode. In this mode, all pixels had the same exposure, which was double the mean pixel exposure of DATA SET B. This led to spurious contributions from events up to $\sim$25~e$^-$ as a sum of two or more events of low energy.

A stable mean value of the SEE was achieved by cleaning the silicon sensor between acquisitions. The average SEE rate over the reactor OFF period from DATA-SET B OFF is 0.11~e$^{-}$/day/pixel. Although stable, this rate is a factor three above the one observed in laboratory conditions with the same system~\cite{Moroni2022}. 

The lack of observable differences between spectra in figure \ref{fig:spectraON2022OFF2022}, along with the impact of adding a cleaning image between acquisitions visualized in the inset of figure~\ref{fig:spectraOFF2022OFF2023}, support the expectation of a competitive reactor ON minus reactor OFF spectrum (signal) from DATA SET B. 

\section{Outlook and Conclusions}

A Skipper-CCD, a low-energy threshold silicon detector, was positioned just 12 meters from the core center of a 2~GWth nuclear reactor power plant. This task presented challenges due to work and security regulations inside the dome, and we succeeded thanks to the close collaboration of the plant's staff members. The performance of this system throughout its initial eighteen months of operation was analyzed.

Although in the current stage, the sensor is provided with 5~cm of lead and $\sim$5~cm of polyethylene shield, no significant differences were detected between the spectra with the reactor ON and OFF down to 120~eV. After the upgrade of the setup and the acquisition protocol, the background spectrum was observed flat down to 45~eV during the last reactor OFF period, with a mean rate of $\sim$35~kdru.

As a first improvement, we recently added a 50 mm-high lead cylinder on top of the copper case of the sensor (numbered as 2 in figure~\ref{fig:system-in-Atucha}, right). The impact of having completed the shielding in the entire 4$\pi$ solid angle will be assessed in the coming months.
Furthermore, in the immediate future, an additional 1300~kg of lead will be added to the setup, enabling a doubling of the shielding thickness.
Thus, we expect to reduce the pile-up and background levels by mitigating the gamma rays responsible for Compton scattering that contribute to the lower energy background, ultimately leading to increased efficiency in event reconstruction.
Increasing the number of sensors is also under consideration, and there is an ongoing development of a new sensor design to scale the mass by a factor of 20.

\begin{figure}[t]
\centering 
\includegraphics[width=0.9\textwidth]
{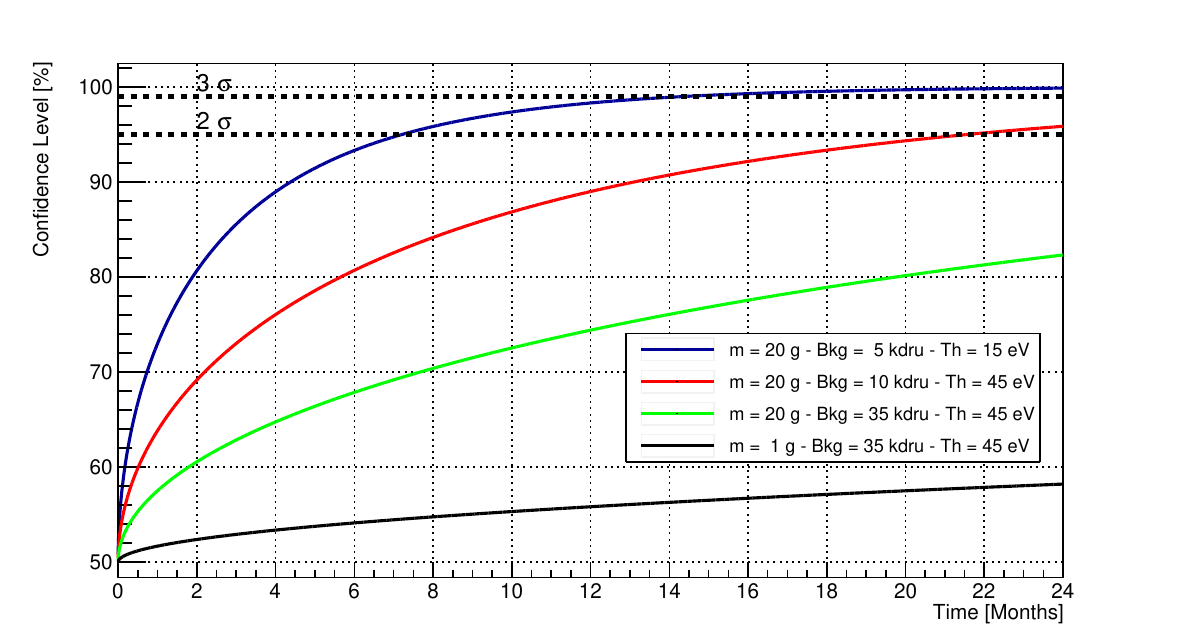}
\caption{Confidence levels as a function of time for the current setup (black line) and three upcoming scenarios: increasing mass up to 20 g (all colored lines), decreasing background down to 10 kdru (red line), and reducing threshold down to 15 eV and background to 5 kdru (blue line).}
\label{fig:outlook}
\end{figure}

Figure~\ref{fig:outlook} depicts the sensitivity curves for this experiment to observe CEvNS as a function of time. The number of events expected by CEvNS was calculated from the indicated threshold (Th = 15~eV and Th = 45~eV) up to 250 keV. The black line corresponds to the sensitivity achieved with the setup described in this work, while the colored lines visualize the impact of increasing mass (green), reducing background (red), and lowering the detection threshold (blue). As can be seen, after increasing mass and improving the background, even without reducing our current threshold, a 90\% confidence level would be achieved in two years. With an additional factor of two in the background and reaching the expected threshold of 15 eV, one and a half years would be enough to reach a 99\% confidence level.

An ongoing reactor ON run is being conducted (DATA SET B, ON in Table~\ref{tab:datasets}). Using this new data, we expect to be able to constrain BSM scenarios that predict a relatively higher rate of events in the eV energy range, such as Light Vector Mediators and milliCharged Particles.
This work contributes to the understanding of the neutrino physics achievable with Skipper-CCD technology inside a nuclear power plant, paving the way for the next generation of kilogram-scale Skipper-CCD short-baseline neutrino experiments~\cite{VIOLETA2021, VIOLETA2022}.

\acknowledgments

We thank the NA-SA team in Argentina for all the support during the deployment and operation of the Skipper-CCD system at Atucha II. This work was supported by Fermilab under DOE Contract No. DE-AC02-07CH11359. We thank Laboratorio de Metrología de Radioisótopos, CAE-CNEA, for providing the HPGe detector and Eneas Kapou from NA-SA for assisting in the transportation of the hard disk with the data outside the plant. We thank Daniel Cartelli from CNEA for creating the artistic plots of the plant and the system. We also want to thank Eduardo Arostegui who insisted on carrying on with the project and facilitated a lot the communication with the plant during the first phase of the deployment.


\bibliographystyle{JHEP}
\bibliography{biblio.bib}

\end{document}